\documentclass[aps,prl,twocolumn,reprint,superscriptaddress,10pt,showpacs]{revtex4-1}
\usepackage{amsmath,amssymb,amsfonts}
\usepackage{graphicx}
\usepackage{txfonts}
\usepackage{color}
\usepackage{math tools}
\usepackage[unicode,breaklinks]{hyperref}
\hypersetup{
    unicode=true,
    a4paper=true,
    plainpages=false,
    pdftitle={Identifying a superfluid Reynolds number via dynamical similarity},
    pdfauthor={M. T. Reeves, T. P. Billam, B. P. Anderson, and A. S. Bradley},
    pdfsubject={Identifying a superfluid Reynolds number via dynamical similarity},
    colorlinks=true,
    linkcolor=blue,
    citecolor=blue,
    filecolor=black,
    urlcolor=blue
}
\newcommand{\eqnreft}[1]{{Eq.~(\ref{#1})}}

\newcommand{\rr}{\mathbf{r}}

\newcommand{\st}{\mathrm{St}}
\newcommand{\re}{\mathrm{Re}}

\begin{document}

\title{Identifying a superfluid Reynolds number via dynamical similarity}

\author{M. T. Reeves}
\thanks{Author to whom correspondence should be addressed}
\email[Email: ]{matt.reeves@postgrad.otago.ac.nz}
\affiliation{Jack Dodd Centre for Quantum Technology, Department of Physics,
University of Otago, Dunedin 9016, New Zealand}
\author{T. P. Billam}
\affiliation{Jack Dodd Centre for Quantum Technology, Department of Physics,
University of Otago, Dunedin 9016, New Zealand}
\affiliation{Joint Quantum Centre (JQC) Durham--Newcastle, Department of  
Physics, Durham University, Durham, DH1 3LE, UK}
\author{B. P. Anderson}
\affiliation{College of Optical Sciences, University of Arizona, Tucson, AZ
85721, USA}
\author{A. S. Bradley}
\thanks{Author to whom correspondence should be addressed}
\email[Email: ]{ashton.bradley@otago.ac.nz}
\affiliation{Jack Dodd Centre for Quantum Technology, Department of Physics,
University of Otago, Dunedin 9016, New Zealand}

\date{\today}


\begin{abstract} The Reynolds number provides a characterization of the
transition to turbulent flow, with wide application in classical fluid
dynamics. Identifying such a parameter in superfluid systems is challenging due
to their fundamentally inviscid nature. Performing a systematic study of
superfluid cylinder wakes in two dimensions, we observe dynamical similarity of
the frequency of vortex shedding by a cylindrical obstacle. The universality of
the turbulent wake dynamics is revealed by expressing shedding frequencies in
terms of an appropriately defined superfluid Reynolds number, ${\rm Re}_s$,
that accounts for the breakdown of superfluid flow through quantum vortex
shedding. For large obstacles, the dimensionless shedding frequency exhibits a
universal form that is well-fitted by a classical empirical relation. In this
regime the transition to turbulence occurs at $\re_s\approx 0.7$, irrespective
of obstacle width. 

\end{abstract}
\maketitle
Turbulence in classical fluid flows emerges from the competition between
viscous and inertial forces. For a flow with characteristic length scale $L$,
velocity $u$, and kinematic viscosity $\nu$, the dimensionless Reynolds number
$\re=uL/\nu$ characterizes the onset and degree of turbulent motion.  A naive
evaluation of the Reynolds number for an ideal superfluid is thwarted by the
absence of kinematic viscosity, suggesting that the classical Reynolds number
of a superfluid is formally undefined~\cite{Barenghi08a,Sasaki2010,Stagg2014}.
However, for sufficiently rapid flows, perfect inviscid flow breaks down and an
effective viscosity emerges dynamically via the nucleation of quantized
vortices~\cite{Frisch92a}. As noted by Onsager~\cite{Onsager:1953va}, the
quantum of circulation of a superfluid vortex, given by the ratio of Planck's
constant to the atomic mass, $h/m$, has the same dimension as $\nu$. This
suggests making the replacement $\nu\rightarrow h/m$, giving a superfluid
Reynolds number $\re_s\sim uLm/h$~\cite{Volovik2003a,Lvov:2014gw}. This
approach is supported by evidence that this quantity accounts for the degree of
superfluid turbulence when $\re_s \gg 1$
\cite{Nore97a,Abid98a,Nore00a,Hanninen07a}, but has yet to be tested by a
detailed study of the transition to turbulence. 

The wake of a cylinder embedded in a uniform flow is a paradigmatic example of
the transition to turbulence \cite{Williamson1996a}, and has been partially
explored in the context of quantum turbulence in atomic Bose-Einstein
condensates (BECs)
\cite{Frisch92a,Winiecki99a,Winiecki00a,Sasaki2010,Stagg2014}. The classical
fluid wakes are \textit{dynamically similar}:  for cylinder diameter $D$ and
free-stream velocity $u$ their physical characteristics are parametrized
entirely by $\re = uD/\nu$. Above a critical Reynolds number, vortices of
alternating circulation shed from the obstacle with characteristic frequency
$f$. As a consequence of dynamical similarity, the associated dimensionless
Strouhal number $\mathrm{St} \equiv fD/u$  takes a universal form when plotted
against the Reynolds number.  In the context of a superfluid, the Strouhal
number is a measurable quantity that can be used to \textit{define} the
superfluid Reynolds number as a dimensionless combination of flow parameters
that reveals dynamical similarity.

In this Letter we numerically study the Strouhal--Reynolds relation across the
transition to turbulence in quantum cylinder wakes of the two-dimensional
Gross-Piteaveskii equation. We develop a numerical approach to gain access to
quasi-steady-state properties of the wake for a wide range of system
parameters, and to accurately determine the Strouhal number $\st$. We find that
plotting $\st$ against a superfluid Reynolds number defined as
\begin{equation}
\re_s \equiv \frac{(u-u_c)D}{\kappa},
\label{SuperfluidReynolds}
\end{equation}
where $u_c$ is the superfluid critical velocity and $\kappa \equiv
\hbar/m$~\footnote{Here choosing $\kappa$ rather than $h/m$ results in a
transition to turbulence near $\mathrm{Re}_s \sim 1$.}, reveals dynamical
similarity in the quantum cylinder wake: for obstacles larger than a few
healing lengths the wakes exhibit a universal $\st$--$\re_s$ relation similar
to the classical form. Furthermore, for these obstacles $\re_s$ characterises
the transition to quantum turbulence, with irregularities spontaneously
developing in the wake when $\re_s \approx 0.7$, irrespective of cylinder size.
 
We consider a Gaussian stirring potential moving at a steady velocity
$\mathbf{u}$ through a superfluid that is otherwise uniform in the $xy$-plane
and subject to tight harmonic confinement in the $z$-direction. In the obstacle
reference frame with coordinate $\rr = \rr_L + \mathbf{u} t$, the time
evolution of the lab-frame wavefunction $\psi(\rr,t) = \psi_L(\rr_L,t)$ is
governed by the Gross-Pitaevskii equation (GPE); \begin{equation} i\hbar
\frac{\partial \psi(\rr ,t)}{\partial t} = (\mathcal{L} - \mathbf{u} \cdot
\mathbf{p} - \mu)\psi(\rr ,t), \end{equation} where $\mu$ is the chemical
potential, $\mathbf{p} = -i\hbar \nabla$, and \begin{equation} \mathcal{L}
\equiv \left[-\frac{\hbar^2 \nabla^2}{2m} + V_s(\rr) +
g_{\rm{2}}|\psi(\rr,t)|^2 \right].  \end{equation}
   \begin{figure}[!t]
\includegraphics[width = \columnwidth]{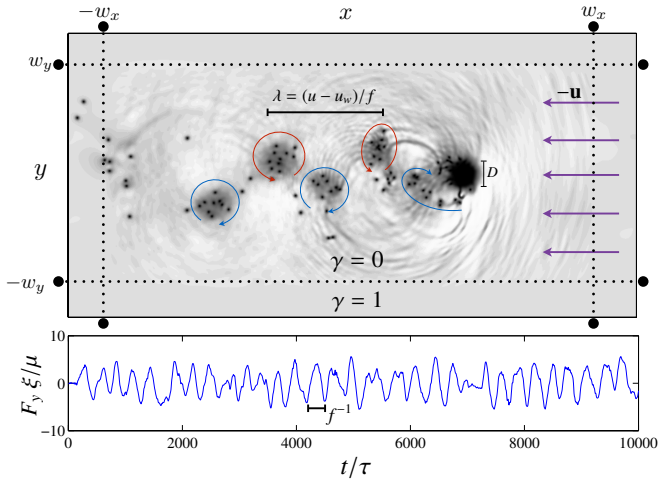}
\caption{(color online)  (Top) A quantum cylinder wake in the quasi-steady
state. Same-sign vortices aggregate into clusters to form a semi-classical
vortex street. Vortices within the fringe region $|x| > w_x$ are unwound in
pairs by imprinting opposite-signed vortices on top of them, thus recycling the
flow to the uniformly translating state as indicated at the right of the
domain. (Bottom) Time series data of the transverse force on the obstacle. The
force exhibits a well-defined frequency, which determines the Strouhal number
for the flow.}
\label{Schematic}
\end{figure}
Here, $g_2 = \sqrt{8\pi}\hbar^2 a_s/ml_z$, where $m$ is the atomic mass, $a_s$
is the $s$-wave scattering length, and  $l_z = \sqrt{\hbar/m\omega_z}$ is the
harmonic oscillator length in the $z$-direction. The trapping in the
$z$-direction is assumed strong enough to suppress excitations along this
direction \footnote{Note that particularly strong confinement is not necessary
to obtain effectively two-dimensional vortex dynamics \cite{Rooney11a}.}. The
stirring potential is of the form $V_s(\rr) = V_0 \exp\{-[(x- x_0)^2
-y^2]/\sigma^2\}$, giving an effective cylinder width, $D = 2a = 2 \sigma
[\log(V_0/\mu)]^{1/2}$, defined by the zero-region of the density in the
Thomas-Fermi approximation. 
In contrast to previous studies~\cite{Sasaki2010,Stagg2014} employing strong
potentials ($V_0 \sim 100\mu$) to approximate a hard-walled obstacle, we use
\textit{soft-walled} obstacles (with $V_0=e\mu$, such that $D=2\sigma$): these
obstacles exhibit a well-defined vanishing-density region, but have a much
lower critical velocity than hard-wall obstacles \cite{Winiecki99a}. A low
critical velocity makes the transition to turbulence --- which must occur
between the critical velocity and the supersonic regime --- more gradual,
aiding our numerical characterization. We find that $D$ gives a good indication
of the effective cylinder width for all obstacles we consider, with vortices
unpinning from the obstacle at $y \approx \pm a$ (see Fig.~\ref{Schematic}).

A key innovation facilitating our study of quasi-steady-state quantum cylinder
wakes is a numerical method to maintain approximately steady inflow-outflow
boundary conditions in the presence of quantum vortices. This method enables us
to evolve cylinder wakes for extremely long times in a smaller spatial domain,
making our numerical experiment computationally feasible. In essence, we extend
the \emph{sponge} or  \emph{fringe method}
\cite{Spalart1989,Colonius2004,Mani2012,Nordstrom99a}, which implements steady
inflow/outflow boundary conditions by ``recycling'' flow in a periodic domain,
to deal with quantum vortices. The spatial region of the numerical simulation
is divided into a ``computational domain'' of interest and a ``fringe domain".
Inside the fringe domain, we use a damped GPE \cite{Tsubota2002,Blakie08a} to
rapidly drive the wavefunction to the lab-frame ground state with chemical
potential $\mu$; a uniform state, free from excitations and moving at velocity
$-\mathbf{u}$ relative to the obstacle, is thus produced at the outer boundary
of the fringe regions. The modified equation of motion is thus
\begin{equation} i\hbar \frac{\partial \psi(\rr,t)}{\partial t} = (\mathcal{L}
- \mathbf{u} \cdot \mathbf{p} - \mu)\psi(\rr,t) -i\gamma(\rr)(\mathcal{L}_f -
\mu)\psi(\rr,t), \label{dGPE} \end{equation}
where the free GPE evolution operator $\mathcal{L}_f \equiv \mathcal{L} -
V_s(\rr)$. At the computational/fringe boundary $(x,y)=(\pm w_x,\pm w_y)$,
$\gamma$ must ramp smoothly from zero to a large value to prevent reflections,
with hyperbolic tangent functions a common choice \cite{Colonius2004}. We set
$\gamma(\rr) = \max[\gamma(x),\gamma(y)],$ where $\gamma(x) = \gamma_0\{ 2 +
\tanh[( x - w_x)/d] - \tanh[(x + w_x)/d]\}/2$ and similarly for $\gamma(y)$. 

Quantum vortices, as topological excitations, decay only at the fluid boundary
or by annihilation with opposite-sign vortices. While damping drives
opposite-signed vortices together at a rate proportional to $\gamma$
\cite{tornkvist_shroder_prl_1997}, relying on this mechanism to avoid vortices
being ``recycled'' around the simulation domain requires a prohibitively large
fringe domain when the wake exhibits clustering of like-sign vortices, a key
feature of the transition to turbulence. Instead, we \emph{unwind}
vortex-antivortex pairs within the fringe domain by phase imprinting an
antivortex-vortex pair on top them, using the rapidly converging expression for
the phase of a vortex dipole in a periodic domain derived in
Ref.~\cite{Billam2013a}. When vortices of only one sign exist within the fringe
region, the same method is used to ``reset'' vortices back near the start of
the fringe ($x = -w_x$) to avoid them being recycled. The high damping in the
fringe domain rapidly absorbs the energy added by this imprinting.

Working in units of the the healing length $\xi = \hbar/\sqrt{m\mu}$, the speed
of sound $c = \sqrt{\mu/m}$ and time unit $\tau = \hbar/\mu$, we discretize  a
spatial domain of $L_x = 512\xi$ by $L_y = 256\xi$ on a grid of $M_x = 1024$ by
$M_y = 512$ points. The obstacle is positioned at $x_0 = 100\xi$, and for the
fringe domain we set $w_x = 220\xi$, $w_y = 100\xi$, $d = 7\xi$ and $\gamma_0
=1$ \footnote{We have verified that the magnitude and frequency of the
transverse force and  the magnitude of the streamwise force on the obstacle are
independent  of the choice of resolution, spatial domain size, details of the
fringe domain, and obstacle location in our simulations. A slightly larger
domain is required for the largest obstacle $D/\xi = 24$ than is quoted in the
main text. For this obstacle, we verify that rescaling $\{L_x,L_y,w_x,w_y,x_0\}
\rightarrow \alpha \{L_x,L_y,w_x,w_y,x_0\}$ (while also scaling $M_x,M_y$ to
maintain the same spatial resolution) yields very similar (within error bars)
Strouhal numbers for $\alpha \approx 1.2$ and $\alpha = 2$.}.

 A typical result from this setup is shown in Fig.~\ref{Schematic}.  We
integrate \eqnreft{dGPE} pseudospectrally, for sufficient time to accurately
resolve the cluster shedding frequency $f$ (see Fig.~\ref{Schematic}, bottom
panel). A small amount of initial noise is added to break the symmetry.
Analyzing obstacles in the range $4 \leq D/\xi \leq 24$ requires integration
times $5000 \leq T/\tau \leq12000$, representing a significant computational
challenge.  To determine the Strouhal number $\st = fD/u$ we calculate the
transverse force on the obstacle from the Ehrenfest relation, $F_y = \int
d^2\rr \; \psi^* (\partial_y V_s) \psi$, with $f$ being defined by the dominant
mode in the frequency power spectrum of $F_y$.  

\begin{figure*}
\includegraphics[width = .9\textwidth]{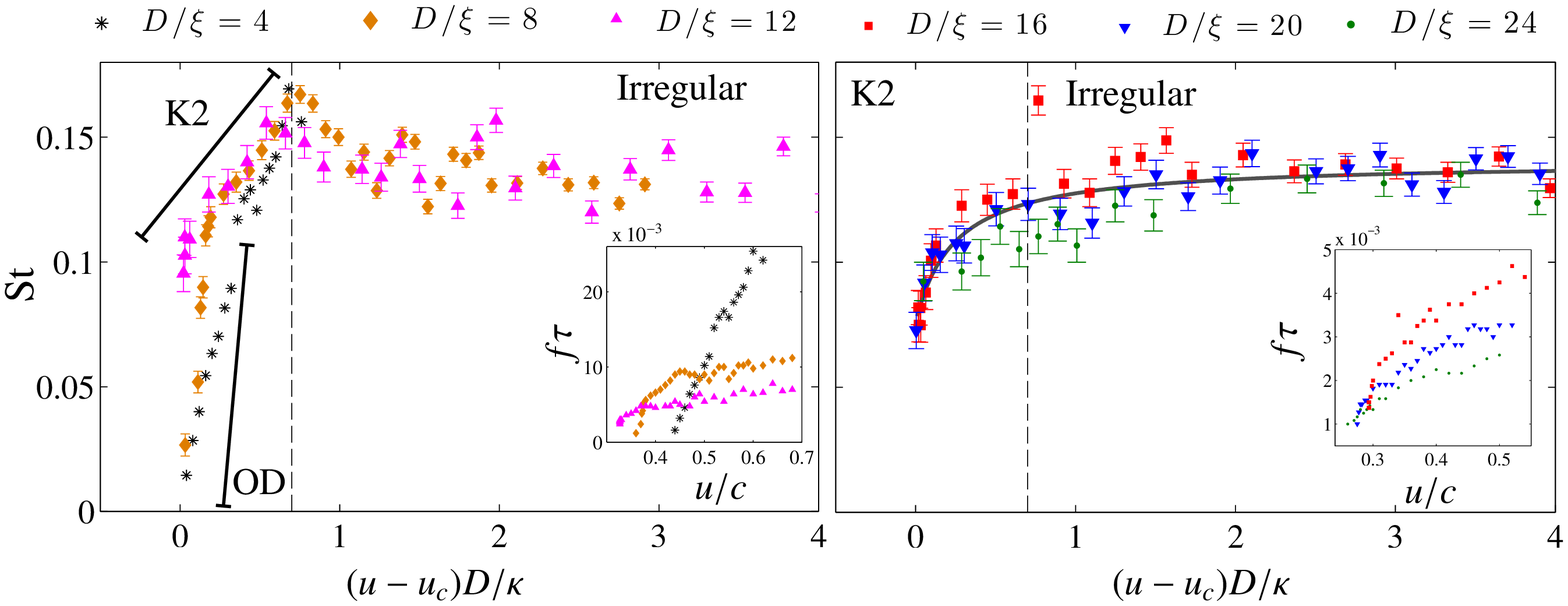}
\caption{ (color online)  Strouhal number plotted as a function of the
superfluid Reynolds number for obstacles of different diameters $D$.  The
dashed lines indicate the transition between regular and irregular wakes (see
text). Solid lines in the left panel indicate regions of oblique dipole (OD)
and charge-2 von K\'arm\'an (K2) shedding. The solid gray line shows the
best-fit curve  $\mathrm{St} = 0.1402 [1 - 0.1126/(\re_s + 0.2456)]$. Error
bars give an indication of the uncertainty in $\st$ due to the Fourier-space
resolution $\Delta f = 1/T$. Insets show the original shedding frequency data
as a function of velocity. The data for $D/\xi = 4$ is truncated as the
shedding frequency becomes poorly defined at higher velocities for this
particular obstacle. In the Supplemental Material \cite{suppinf} we provide
movies showing condensate density and vortex-cluster dynamics in each shedding
regime.}
\label{Strouhals}
\end{figure*}

Our main results are shown in Fig.~\ref{Strouhals}, where the Strouhal number
$\st$ is plotted against the superfluid Reynolds number $\re_s =
(u-u_c)D/\kappa$ for a range of obstacle diameters $D$ [insets show shedding
frequency $f$ against velocity $u$]. In the Supplemental Material
\cite{suppinf} we provide movies showing condensate density and vortex-cluster
dynamics for representative sets of parameters. The obstacles are broadly
classified as quantum ($\sigma \leq 12\xi$, left) or semi-classical ($\sigma >
12\xi$, right).  For quantum obstacles the vortex core size influences the
shedding dynamics, and the $\st$-$\re_s$ curve exhibits three distinct regimes:
At low $\re_s$, vortex dipoles are released obliquely from the obstacle (OD
regime), and St rises sharply with $\re_s$. As $\re_s$ is increased, the
gradient of the $\st$-$\re_s$ curve drops sharply when a charge-2 von
K\'{a}rm\'{a}n vortex street \cite{Sasaki2010} appears (K2 regime).  The
Strouhal number peaks at  $\re_s \sim 0.7, \mathrm{St} \approx 0.16$, and
beyond this point the shedding becomes irregular, and the Strouhal number
gradually decreases towards $\mathrm{St} \approx 0.14$. 
The $\st$-$\re_s$ data conform to a single curve rather well when compared
against the $f$ vs. $u$  data shown in the inset, apart from variation in the
OD regime at low $\re_s$. This can be attributed to the influence of vortex
core structure on shedding, which is most pronounced for $D/\xi =4$. At $D/\xi
= 12$ the curve becomes very steep, and dipole shedding seems to disappear.

For semi-classical obstacles (right panel of Fig.~\ref{Strouhals}), the
$\st$-$\re_s$ curve is qualitatively different. Obstacles with $D/\xi \geq 12$
appear to lack a stable OD regime \footnote{For $D/\xi = 16$, even resolving
the critical velocity to within $\Delta u/c = 2\times10^{-4}$ does not reveal a
clear OD regime.}, and the most steeply-rising region of the $\st$-$\re$ curve
corresponds to the K2 regime.  The peak seen in the  $\st$-$\re_s$ curve for
quantum obstacles is generally absent (with a remnant for $D/\xi = 16$), and
the $\st$-$\re_s$ data conform to a universal curve extremely well for $\re_s
\lesssim 0.5$ and $\re_s \gtrsim 2$, and to a lesser extent around $\re_s = 1$.
This discrepancy may be an effect of using a soft-walled obstacle, for which
varying $\sigma$ for fixed $V_0$ leads to a slight change in the density
profile near the obstacle.  Remarkably, the $\st$-$\re_s$ curve for the
semiclassical obstacles is well-fitted by the  formula $\st = \st_\infty[1 -
\alpha/(\re_s + \beta)]$~ \footnote{ The need for the shift $\beta$ in the fit
shown in Fig.~\ref{Strouhals} is a consequence of the fact that the vortex
street in a classical fluid does not appear until $\re \gtrsim 40$, whereas for
our semiclassical obstacles it emerges immediately above $u_c$ (i.e., for
$\re_s>0$).}, which is similar to the classical form $\st = \st_\infty(1 - A/
\re)$~ \cite{Roshko1954}.

\begin{figure}
\includegraphics[width = \columnwidth]{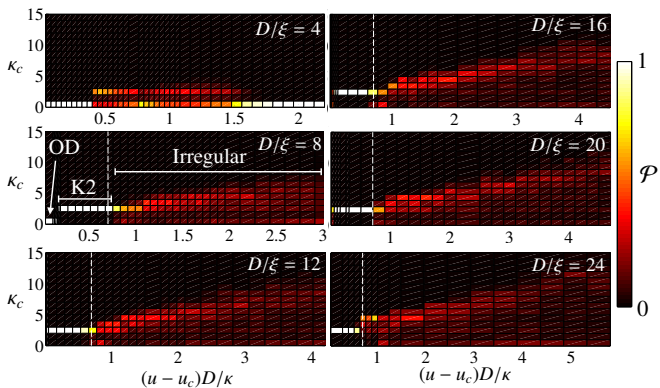}
\caption{Cluster charge probability distribution $\mathcal{P}(\kappa_c,\re_s)$,
which shows the probability that a vortex belongs to a cluster of charge
magnitude $\kappa_c$. Note that $\kappa_c=0$ corresponds to a dipole and
$\kappa_c=1$ corresponds to a free vortex. The vertical dashed line shows the
value $\re_s = 0.7$ at which the probability distribution suddenly spreads,
indicating that the wake has developed irregularities. The three different
shedding regimes observed are labelled for the case $D/\xi = 8$.}
\label{PvsRe}
\end{figure}

To test whether $\re_s$ provides an accurate indicator of the transition to
quantum turbulence, in Fig.~\ref{PvsRe} we show the vortex-cluster charge
probability distribution, $\mathcal{P}(\kappa_c,\re_s)$. This indicates the
probability of any vortex belonging to a cluster of charge $\kappa_c$, as
determined by the recursive cluster algorithm of
Ref.~\cite{reeves_etal_prl_2013}. The transition to turbulence manifests as an
abrupt spreading in $\mathcal{P}$ at $\re_s \approx 0.7$. The distribution
$\mathcal{P}$ is similar for all obstacles except the smallest ($D/\xi=4$)
where high $\re_s$ vortex turbulence is suppressed by compressible effects due
to the transsonic velocities involved. Notice that the distribution is close to
independent of obstacle size for larger obstacles ($D\geq 12\xi$). We find that
the K2 regime persists for a significant range of $\re_s$ even for large $D$,
in contrast to Ref.~\cite{Sasaki2010}. We suggest the vanishing of the K2
regime at large $D$ seen in Ref.~\cite{Sasaki2010} may be due to the higher
critical velocity of the hard-walled obstacle. We find no regular
charge-$\kappa_c$ von K\'{a}rm\'{a}n regimes (K$\kappa_c$ regimes) other than
K2. The lack of a K1 regime, the focus of von K\'{a}rm\'{a}n's original
analysis of vortex streets \cite{Saffman1995}, suggests that the additional
degree of freedom provided by the internal length scale of the charge-2 cluster
is what enables stable vortex shedding in the K2 regime. The lack of
K$\kappa_c$ regimes for $\kappa_c>2$ appears to be due to instabilities;
although regimes do exist where $\mathcal{P}$ is strongly peaked around
$|\kappa_c | > 2$, such regimes do not appear to be stable.  

The superfluid Reynolds number $\re_s$ introduced in
\eqnreft{SuperfluidReynolds} serves as a good control parameter for the
transition to turbulence, which occurs at $\re_s \approx 0.7$ for all obstacle
sizes investigated except $D/\xi = 4$ (it is expected to fail for $D
\rightarrow \xi$, where dynamical similarity is lost).  The definition of
$\re_s$ in terms of $u-u_c$ is intuitively appealing: the subtraction of $u_c$
becomes unimportant in the classical limit (where $u_c$ vanishes) and when
$\re_s \gg 1$, consistent with previous observations
\cite{Nore97a,Abid98a,Nore00a}.  Subtracting $u_c$  is consistent with previous
arguments that corrections to the Reynolds number formula are necessary for
quantum obstacles~\cite{Hanninen07a}, and reflects the fact that in a pure
superfluid an effective viscosity due to quantum vortices is only ``activated''
once vortices are nucleated. 

Although $\re_s$ takes on small values here compared to the Reynolds number of
classical cylinder wakes,  we note the close correspondence between the
$\st$-$\re_s$ curve obtained here and the classical $\st$-$\re$ curve. The
latter rises steeply when the shedding is regular, and reaches a plateau as the
shedding becomes irregular \cite{Roshko1954}. This correspondence suggests that
$\mathrm{Re}_s \sim 0.7$ may be roughly equivalent to $\mathrm{Re} \sim 200$.
The fact that the $\st$-$\re_s$ curves approach a universal form for different
obstacle sizes suggests that the wake structure is insensitive to considerable
changes in Mach number, which occur between different obstacle widths at fixed
$\re_s$, consistent with the observation that the wake is dominated by vortex
shedding even into the transonic regime \cite{Winiecki00a}.  The discrepancy
between the asymptotic values of $\st$ found here and in the classical case
appears to be mainly due to the use of soft-walled obstacles: we have confirmed
that simulations with $V_0/\mu = 10\exp(1)$ and $D/\xi = 20$ produce a
qualitatively similar $\st$-$\re_s$ curve to Fig.~\ref{Strouhals} but with
higher asymptote $\st_\infty\approx 0.16$. For the hard-wall obstacle
\footnote{$V_0/\mu = 100$, $u/c = 0.51659$, $\sigma/\xi = 1.5811$ ($D/\xi =
6.7861$)}  of Ref. \cite{Sasaki2010} we find $\st \approx 0.18$ for velocities
that give a vortex street, in reasonable agreement with classical observations
where $\st_\infty\approx 0.2$~\cite{ahlborn2002,Roshko1954b}. The lower
Strouhal number of the soft-walled obstacle suggests that it  is ``bluffer''
than the hard-walled one, in the sense that it produces a wider wake for a
given obstacle dimension $D$ \cite{Roshko1954b}.

The K2 regime should be accessible to current BEC
experiments~\cite{Sasaki2010}, since the wake is stable and easily identified.
Accessing the high $\re_s$ regime with fine resolution may be experimentally
challenging, however, the low $\re_s$ turbulent regime, particularly near the
transition, should be accessible in current BEC experiments.  In this regime
the Strouhal number should be measurable, since the induced wake velocity  $u_w
\rightarrow 0$ \cite{Roshko1954} and thus the average streamwise cluster
spacing $\lambda = (u-u_w)/f \rightarrow u/f$ determines $\st = D/\lambda$. 

In conclusion, we have developed a vortex-unwinding fringe method to study
quasi-steady-state quantum cylinder wakes, revealing a superfluid Reynolds
number $\re_s$ that controls the transition to turbulence in the wake of an
obstacle in a planar quantum fluid. The expression for $\re_s$ resembles the
classical form, modified to account for the critical velocity at which
effective superfluid viscosity emerges. As the critical velocity encodes
details of geometry and the microscopic nature of the superfluid, the general
form of $\re_s$ suggests that it may apply to a broad range of systems, much
like the classical Reynolds number. We thus conjecture that our work may
provide a useful characterisation of turbulence in any superfluid, such as
liquid helium~\cite{Bewley08b}, polariton condensates~\cite{Tosi:2011kv}, and
BEC-BCS superfluidity in Fermi gases~\cite{Zwierlein:2005kl}.

\acknowledgments
We thank A.~L.~Fetter for bringing Ref.~\cite{Onsager:1953va} to our attention.
We acknowledge support from The New Zealand Marsden Fund and a Rutherford
Discovery Fellowship of the Royal Society of New Zealand (ASB), and the
University of Otago (MTR). TPB was partly supported by the UK EPSRC
(EP/K030558/1). BPA was supported by the US National Science Foundation
(PHY-1205713).

\end{document}